\documentclass{pnastwo}
\usepackage[dvips]{graphicx}
\usepackage{amssymb,amsfonts,amsmath}
\contributor{Submitted to Proceedings
of the National Academy of Sciences of the United States of America}
\url{www.pnas.org/cgi/doi/10.1073/pnas.0709640104}
\copyrightyear{2008}
\issuedate{Issue Date}
\volume{Volume}
\issuenumber{Issue Number}

\begin{document}


\title{The role of random electric fields in relaxors}

\author{Daniel Phelan\affil{1}{NIST Center for Neutron Research, National Institute of Standards and Technology, Gaithersburg, Maryland 20899, USA}, Christopher Stock\affil{1}{}\affil{2}{School of Physics and Astronomy, University of Edinburgh, Edinburgh, UK EH9 3JZ}, Jose A. Rodriguez-Rivera\affil{1}{}\affil{3}{Department of Materials Science and Engineering, University of Maryland, College Park, Maryland 20742, USA}, Songxue Chi\affil{1}{}\affil{3}{}, Juscelino Le\~{a}o\affil{1}{}, Xifa Long\affil{4}{Department of Chemistry and 4D LABS, Simon Fraser University, Burnaby, BC, V5A 1S6, Canada}, Yujuan Xie\affil{4}{}, Alexei A. Bokov\affil{4}{}, Zuo-Guang Ye\affil{4}{}, Panchapakesan Ganesh\affil{5}{Center for Nanophase Materials Sciences, Oak Ridge National Laboratory, Oak Ridge, Tennessee 37831-6494, USA} \and Peter M. Gehring\affil{1}{}\affil{6}{To Whom Correspondence Should Be Addressed (Email: peter.gehring@nist.gov)}}

\contributor{Submitted to Proceedings of the National Academy of Sciences of the United States of America}

\maketitle

\begin{article}

\begin{abstract}
PbZr$_{1-x}$Ti$_x$O$_3$ (PZT) and Pb(Mg$_{1/3}$Nb$_{2/3}$)$_{1-x}$Ti$_x$O$_3$ (PMN-$x$PT) are complex lead-oxide perovskites that display exceptional piezoelectric properties for pseudorhombohedral compositions near a tetragonal phase boundary. In PZT these compositions are ferroelectrics, but in PMN-$x$PT they are relaxors because the dielectric permittivity is frequency dependent and exhibits non-Arrhenius behavior.  We show that the nanoscale structure unique to PMN-$x$PT and other lead-oxide perovskite relaxors is absent in PZT and correlates with a greater than 100\% enhancement of the longitudinal piezoelectric coefficient in PMN-$x$PT relative to that in PZT. By comparing dielectric, structural, lattice dynamical, and piezoelectric measurements on PZT and PMN-$x$PT, two nearly identical compounds that represent weak and strong random electric field limits, we show that quenched (static) random fields establish the relaxor phase and identify the order parameter.
\end{abstract}


\section*{Significance Statement}
Relaxors are characterized by a frequency-dependent peak in the dielectric permittivity and are critical to modern technological applications because they exhibit large dielectric constants and unparalleled piezoelectric coefficients.  Despite decades of study a fundamental understanding of the origin of relaxor behavior is lacking.  Here we compare the structural, dynamical, dielectric, and piezoelectric properties of two highly similar piezoelectric lead-oxide materials: ferroelectric PbZr$_{1-x}$Ti$_x$O$_3$ and relaxor Pb(Mg$_{1/3}$Nb$_{2/3}$)$_{1-x}$Ti$_x$O$_3$.  Random electric fields are implicated as the genesis of relaxor behavior, and the diffuse scattering associated with short-range polar order is identified as the order parameter.  The piezoelectric response is found to be greatly amplified in crystals that display this diffuse scattering.

\section*{Introduction}

\subsection{Relaxors and random electric fields}
\ The remarkable electromechanical properties of lead-oxide, perovskite ($AB$O$_3$) relaxors such as Pb(Mg$_{1/3}$Nb$_{2/3}$)$_{1-x}$Ti$_x$O$_3$ (PMN-$x$PT) and Pb(Zn$_{1/3}$Nb$_{2/3}$)$_{1-x}$Ti$_x$O$_3$ (PZN-$x$PT) have inspired numerous attempts to understand the piezoelectricity in terms of the structural phase diagram, which contains a steep morphotropic phase boundary (MPB) separating pseudo-rhombohedral and tetragonal states over a narrow compositional range where the piezoelectricity is maximal.~\cite{Bellaiche,Fu,Jin,Kutnjak,Xu}  These materials exhibit very low strain-electric field hysteresis, extremely large dielectric constants, and record-setting piezoelectric coefficients at room temperature that form an unusually appealing set of properties with the potential to revolutionize a myriad of important technological applications spanning medical diagnostic sonography, military sonar, energy harvesting, and high-precision actuators.~\cite{Park,Uchino,Kandilian}  Many researchers have argued that quenched random electric fields (REFs) play a central role in establishing the relaxor phase, in part because the $B$-sites of all known lead-oxide perovskite relaxors are occupied by random mixtures of heterovalent cations.~\cite{Vugmeister,Westphal,Stock,Tinte,Cowley}  However there is ample theoretical work that suggests relaxor behavior can occur in the absence of REFs.~\cite{Pirc,Grinberg_REF,Akbarzadeh_2012}  In fact it has not been proven conclusively that REFs are essential to the relaxor state or that they play any role in the ultrahigh piezoelectricity.  These basic questions persist in the face of decades of research mainly because there exists no rigorous definition of what a relaxor is, i.\ e.\ there is no precise mathematical formulation of the relaxor order parameter.  To date any material for which the real part of the dielectric permittivity $\epsilon'(\omega,T)$ exhibits a broad peak at a temperature $T_{max}$ that depends strongly (and in some cases only weakly) on the measuring frequency, $\omega$, is classified as a relaxor. This definition has been applied equally to PMN and PZN, which possess strong REFs, as well as to specific compositions of K(Ta$_{1-x}$Nb$_x$)O$_3$ (KTN), (K$_{1-x}$Li$_x$)TiO$_3$ (KLT), Ba(Zr$_{1-x}$Ti$_x$)O$_3$ (BZT), and Ba(Sn$_{1-x}$Ti$_x$)O$_3$ (BST),~\cite{Samara_Review,Liu_2007} which are all composed of homovalent cations and thus have much weaker REFs.  The REFs in these latter materials are believed to result (primarily) from cation off-centering and are non-zero when this off-centering is static.  We will use the term ``weak REF limit" to make clear that the REFs in such materials are not necessarily zero.

To identify REF-specific properties uniquely one must compare identical systems that differ only in the strength of the REFs.  While such an idealized situation does not exist, the recent breakthrough in the growth of millimeter-size, high-quality, single crystals of Pb(Zr$_{1-x}$Ti$_x$)O$_3$ (PZT)~\cite{YePZT} has finally provided experimentalists with a nearly perfect model system with which such a comparison can be made to classic relaxor systems such as PMN-$x$PT.  Both PZT and PMN-$x$PT are chemically-disordered, lead-oxide perovskites, and both exhibit similar phase diagrams as a function of Ti concentration that contain an MPB near which exceedingly large piezoelectric coefficients are observed.  Moreover, both materials possess nearly identical average $B$-site ionic radii; the average $B$-site ionic radius of PMN is equal to that of PZT with a composition of 46\% Ti.  But whereas PMN possesses strong REFs, PZT is composed of homovalent Zr$^{4+}$ and Ti$^{4+}$ $B$-site cations and thus is representative of the weak REF limit.  In this article we present a combination of dielectric permittivity, piezoelectric, and neutron scattering measurements on each material that indicate that strong REFs are necessary to the formation of the relaxor phase in lead-oxide perovskites and greatly amplify the piezoelectric response.  We develop a heuristic model to motivate the order parameter associated with the relaxor phase that suggests a modified definition of relaxors.

Fig.~1a illustrates how the real part of the dielectric permittivity of PMN varies with temperature at four different measuring frequencies.  From 10$^2$\,Hz to 10$^5$\,Hz the position of the maximum dielectric permittivity shifts to higher temperature by 18\,K; this behavior is the hallmark of all relaxors.  By contrast, the same measurement made over the same frequency range on a single crystal specimen of PZT with $x=0.325$ is shown in Fig.~1b, and no shift with temperature is evident.  Instead the dielectric response of PZT is consistent with that of a conventional ferroelectric phase transition near 600\,K.  Thus, despite the many similarities between PZT and PMN, PZT is not a relaxor.  This result is consistent with the theoretical work of Grinberg \emph{et al.} who used density functional theory to parameterize a phenomenological Landau model of the dielectric frequency dispersion in terms of the average $B$-site cation displacement from the high-symmetry cubic structure, $\overline{D}_B$, and the second moment of the valence of the two $B$-site-cation nearest neighbors of each oxygen atom, $\langle V^2 \rangle$, which is a measure of the REF strength.~\cite{Grinberg_REF}  Using only these two input parameters, the Grinberg model yields remarkable quantitative agreement with all of the 23 lead-based, perovskite relaxor systems for which both the local cation order and dielectric response have been measured.  Interestingly, this model permits a frequency-dependent dielectric response even for materials where $\langle V^2 \rangle = 0$ (i.e. in the weak REF limit) provided that the average $B$-site cation displacement is not too large ($\overline{D}_B < 0.11$\,\AA).  In the case of PZT there is no frequency dependence because the average $B$-site cation displacement $\overline{D}_B$ is large enough (much larger than that in PMN) to stabilize long-range ferroelectric order.  But there exist other disordered perovskite materials composed of homovalent $B$-site cations for which $\langle V^2 \rangle = 0$ and $\overline{D}_B$ is small that do exhibit a strongly frequency-dependent dielectric response.  Examples include compositions of BZT ($0.25 \le x \le 0.42$) and BST ($0.20 \le x \le 0.50$),~\cite{Liu_2007} for which an average cubic symmetry is retained at all temperatures studied.  Here we wish to consider the following fundamental questions: (1) are quenched REFs essential to the relaxor phase observed in the lead oxides, (2) what is the relaxor order parameter, (3) is there a dynamical signature of the relaxor phase analogous to the soft mode that characterizes displacive ferroelectrics, and (4) do REFs play a role in the ultrahigh piezoelectric response?

\begin{figure}[t]
\centering
\includegraphics[width=90mm]{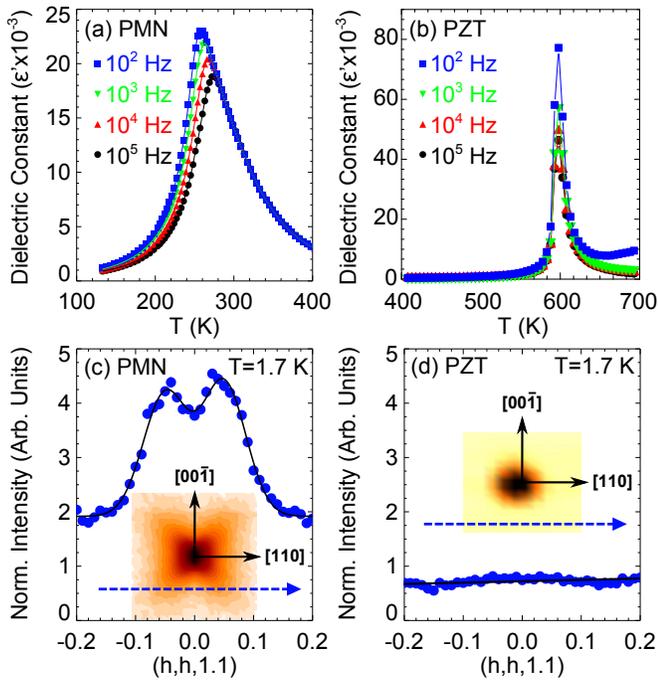}
\caption{Temperature dependence of the real part of the dielectric permittivity measured along [001] at 10$^2$\,Hz, 10$^3$\,Hz, 10$^4$\,Hz, and 10$^5$\,Hz for (A) PMN and (B) PZT ($x=0.325$).  Neutron elastic diffuse scattering intensity (solid blue dots) measured using the BT7 thermal-neutron spectrometer at 1.7\,K along [110] in the $(HHL)$ scattering plane for (C) PMN and (D) PZT.  The strong, diffuse scattering peaks shown in panel (C) for PMN result from the large out-of-plane $\vec{Q}$ resolution, which overlaps with the four rods of diffuse scattering that point out of the $(HHL)$ scattering plane and are centered at (001).  These peaks are absent for PZT as shown in panel (D).  The insets to panels (C) and (D) illustrate the shape of constant elastic-scattering-intensity contours measured near (001) in the $(HHL)$ scattering plane for the same PMN and PZT crystals at 300\,K and 10\,K, respectively, and are plotted on a log-intensity scale.  Further details are provided in the SI Materials and Methods.}
\end{figure}

\section*{Results}

\subsection{Static signature: short-range polar order}
\ To answer these questions we compared the nanoscale structures of PMN and PZT using neutron-based measurements of the elastic diffuse scattering, which is arguably the single most interesting and intensely examined property of lead-oxide relaxors to date.  Strong, temperature-dependent, and highly anisotropic diffuse scattering has been documented in many different lead-oxide, perovskite relaxors using x-ray and neutron-scattering methods.~\cite{Vakhrushev_1995,You_1997,Malibert_1997,Hirota_2002,Gvasaliya_2003,Xu_2004,Ohwada_2006,Mihailova_2008}  The inset to Fig.~1c illustrates the anisotropic nature of this diffuse scattering in PMN via a log-scale intensity map of the well-known, butterfly-shaped diffuse scattering contours that decorate the (001) Bragg peak in the $(HHL)$ scattering plane defined by the pseudocubic [110] and [001] crystallographic axes.  We measured the diffuse scattering intensity at very low temperature (1.7 \,K), where it is strongest, by scanning the neutron momentum transfer $\vec{Q}$ along the trajectory defined by the dashed blue arrow in Fig.~1c.  These data are given by the solid blue dots, and two strong peaks are evident where this trajectory clips the two lower wings of the butterfly-shaped contours.  The origin of this diffuse scattering is extremely controversial.  Numerous models of the underlying short-range structural order have been proposed to explain the experimental data.~\cite{Xu_2004,Welberry_2006,Pasciak_2007,Ganesh_2010,Cervellino_2011,Bosak_2011}  Neutron backscattering measurements,~\cite{Gehring_1998,Meyer_2003} which provide very sharp energy resolution ($\Delta E \sim 1$\,$\mu$eV), show that truly elastic (i.\ e.\ static) diffuse scattering in PMN first appears at a temperature $T_d=420 \pm 20$\,K that coincides with that at which the lowest-frequency transverse optic (TO) phonon is softest and overdamped; it also grows monotonically on cooling.~\cite{Gehring_2009}  A monotonic increase of the diffuse scattering on cooling is observed in all pseudo-rhombohedral compositions of PMN-$x$PT and PZN-$x$PT.~\cite{Matsuura_2006,Xu_2004}  Other neutron and x-ray studies have demonstrated that relaxor diffuse scattering is strongly affected by external electric fields~\cite{Gehring_2004,Stock_2007} and vanishes at high pressures.~\cite{Chaabane_2003,Mihailova_2008} For these reasons many researchers identify the diffuse scattering with static, nanometer-scale regions of short-range polar order known as polar nanoregions that condense from the soft TO mode on cooling from high temperatures well above $T_{max}$.~\cite{Hirota_2002,Ganesh_2010}  We emphasize that all lead-oxide, perovskite relaxors for which single crystals have been examined with neutron or x-ray scattering techniques exhibit similar anisotropic, temperature-dependent diffuse scattering contours including PMN, PZN, Pb(Sc$_{1/2}$Nb$_{1/2}$)O$_3$, Pb(Mg$_{1/3}$Ta$_{2/3}$)O$_3$, Pb(In$_{1/2}$Nb$_{1/2}$)O$_3$, and Pb(Sc$_{1/2}$Ta$_{1/2}$)O$_3$.~\cite{Vakhrushev_1995,Xu_2004,Malibert_1997,Ganesh_2010,Gvasaliya_2003,Cervellino_2011,Ohwada_2006,Mihailova_2008}  All of these compounds are composed of heterovalent $B$-site cations; hence $\langle V^2 \rangle > 0$.  In addition, a recent neutron-scattering study discovered that the lead-free relaxor (Na$_{1/2}$Bi$_{1/2}$)TiO$_3$ (NBT), an $A$-site disordered perovskite composed of heterovalent Na$^+$ and Bi$^{3+}$ cations, displays static diffuse scattering contours characterized by a temperature-dependence and anisotropy like that found in PMN.~\cite{Ge_2013}  This suggests that the physics that governs this diffuse scattering is not necessarily limited to lead-based, perovskite relaxors.

By contrasting the elastic diffuse scattering from PMN to that from PZT, we can isolate the influence of quenched REFs on the nanometer-scale structure of lead-oxide, perovskite relaxors.  To this end we examined the neutron diffuse scattering from a single crystal of PZT with a Ti concentration $x = 0.325$ under exactly the same conditions as those used to measure the diffuse scattering from the PMN single crystal.  To account precisely for differences in crystal size and beam illumination, the diffuse scattering data from both crystals were normalized by the integrated intensity of a transverse acoustic (TA) phonon also measured under identical conditions.  The resulting data, given by the solid blue dots in Fig.~1d, shows that the diffuse scattering for the non-relaxor PZT is at least a factor of 20 weaker than that in PMN.  The inset to Fig.~1d shows a log-scale intensity map of the elastic scattering from the same PZT crystal at very low temperature (10\,K) in the $(HHL)$ scattering plane near (001), and no evidence of diffuse scattering is present.  Additional data measured on PZT in the $(HK0)$ scattering plane at both low (16\,K) and high (620\,K) temperatures confirm this finding (see Fig.~S1 in the SI Appendix). Static diffuse scattering was recently reported in a single crystal of PZT with $x = 0.475$ using x-ray inelastic scattering techniques.~\cite{Burkovsky}  However these data were obtained with an energy resolution nearly 10 times coarser than ours; hence there is a possibility of phonon contamination.  This static diffuse scattering was modeled using a Huang formalism, which is inconsistent with that seen in relaxors, and was attributed to elastic deformations induced by defect centers of tetragonal symmetry.  This is not unexpected for a composition so close to the MPB.  Indeed such Huang scattering would have impaired our search for relaxor-like diffuse scattering except for the fact that our PZT composition ($x = 0.325$) lies in the purely pseudo-rhombohedral regime.

As the only significant difference between PMN and PZT lies in the huge disparity in REF strength, our neutron data establish a direct causal link between strong REFs and the temperature-dependent, anisotropic diffuse scattering in lead-oxide, perovskite relaxors.  In the same vein, our dielectric permittivity measurements demonstrate that strong REFs are responsible for the relaxor-like behavior.  Together these results imply the existence of a fundamental connection between the relaxor phase and the anisotropic, temperature-dependent diffuse scattering.  We motivate this picture through a heuristic model, developed in the SI Materials and Methods, that allows us to relate the strength of the REFs to the diffuse scattering intensity.  We thus identify the relaxor order parameter with the static, anisotropic diffuse scattering associated with short-range polar order, but only in the sense that it marks the transition temperature from the high-temperature paraelectric phase (where it is zero) to the low-temperature relaxor state (where it is non-zero).  Indeed, as noted by Cowley \emph{et al.}~\cite{Cowley} and Prosandeev \emph{et al.}~\cite{Prosandeev} the strictly elastic diffuse scattering in PMN increases on cooling in a manner similar to that of an order parameter at a conventional phase transition, and this also holds true for all pseudo-rhombohedral compositions of PMN-$x$PT and PZN-$x$PT.  As first discussed by Westphal \emph{et al.}~\cite{Westphal} and later Stock \emph{et al.},~\cite{Stock} this identification is consistent with the ideas of Imry and Ma who showed that magnetic systems with continuous symmetry will break up into short-range ordered domains in the presence of random magnetic fields.~\cite{Imry_1975}  In relaxors like PMN, long-range ferroelectric order is stifled by the formation of analogous polar domains on cooling.  By contrast, our data show that no such domains form in ferroelectric PZT because the REFs are too weak.  This picture naturally explains why the diffuse scattering in relaxors vanishes for compositions on the tetragonal side of the MPB; the uniaxial anisotropy there stabilizes the ferroelectric state.

\begin{figure}[t]
\centering
\includegraphics[width=85mm]{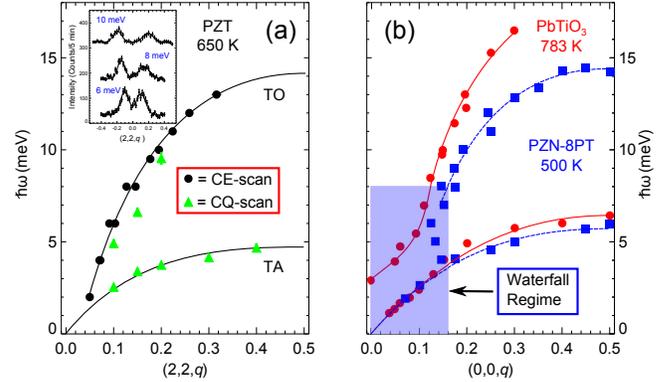}
\caption{Dispersions of the TA and soft TO modes measured in the $(HHL)$ scattering plane near (220) in the cubic phases of (a) PZT and (b) PbTiO$_3$ (Ref. \cite{Shirane_1970}) and PZN-8\%PT (Ref. \cite{Gehring_waterfall}).  The green triangles/black circles in panel (a) denote the locations of maximum scattering intensity in constant-$\vec{Q}$/constant-$\hbar\omega$ scans, respectively; the inset shows constant-$\hbar\omega$ scans measured at 650\,K at 6, 8, and 10\,meV (data have been displaced vertically for clarity).  All lines are guides to the eye.}
\end{figure}

\subsection{Dynamic signature:  the Waterfall effect versus zone-boundary soft modes}
\ We next compare the lattice dynamical properties of PZT and PMN to look for a dynamic signature of the relaxor phase.  PMN exhibits several anomalous dynamical features that are also seen in other relaxor compounds such as PZN-$x$PT and Pb(Mg$_{1/3}$Ta$_{2/3}$)O$_3$; thus the availability of a high quality PZT single crystal provides us with a rare opportunity to determine if any of these are specific to lead-based perovskite relaxors.  In particular, the diffuse scattering in PMN, which is purely relaxational above $T_d = 420$\,K,~\cite{Stock_NSE} has profound effects on the lifetimes of low-frequency TA modes.~\cite{Stock_2005,Stock_2012}  The best-known dynamic anomaly is the ``waterfall effect," a term coined to characterize the false impression that the soft TO phonon branch in PZN-8\%PT drops steeply into the TA branch at a non-zero wave vector $|\vec{q}| \approx 0.2$\,\AA$^{-1}$.~\cite{Gehring_waterfall}  The waterfall effect results from a strong damping of the TO mode that occurs at significantly larger wave vectors than happens in cubic PbTiO$_3$.~\cite{Shirane_1970}  The cause of this damping is also controversial.  Model simulations based on standard TO-TA mode coupling theory provide qualitative agreement with experimental data measured on PZN-8\%PT.~\cite{Hlinka_waterfall}  However TO-TA mode coupling has been shown to be weak in PMN and therefore unlikely to be the origin of the waterfall effect.~\cite{Stock_2005,Stock_2012}  An alternative explanation is that the waterfall effect arises from disorder introduced through the heterovalent nature of the $B$-site cations, i.\ e.\ random dipolar fields.~\cite{Stock_2012}  If this idea is correct, then the waterfall effect should be absent in PZT or qualitatively different from that observed in PMN and other relaxors.  Another unusual dynamical feature is the simultaneous appearance of highly-damped zone-boundary soft modes at the R and M-point Brillouin zone boundaries in PMN,~\cite{Swainson_2009} which exhibit a temperature dependence that tracks that of the soft, zone-center TO mode, and which give rise to broad superlattice peaks at the same locations.  This behavior indicates that competing antiferroelectric and ferroelectric fluctuations are present in PMN, a situation that has been applied theoretically to SrTiO$_3$.~\cite{Zhong_1995}  It was recently shown that the waterfall effect is present in PMN-60\%PT, a composition that lies well beyond the MPB.~\cite{Stock_2006}  Because PMN-60\%PT exhibits no diffuse scattering and no discernable frequency dependence to the dielectric permittivity, the waterfall effect cannot be associated with the relaxor phase.  The observation of a waterfall effect in PZT would provide additional evidence for this conclusion.  By contrast, the soft and highly-damped R and M-point zone boundary modes seen in PMN are absent in PMN-60\%PT ~\cite{Stock_2006} but have been observed in the relaxor PZN-5.5\%PT; thus these broad modes may constitute a genuine dynamical signature of relaxors.

To determine if these dynamical anomalies are intrinsic to the relaxor state, we present the first neutron inelastic scattering data on PZT, which were obtained in the $(HHL)$ scattering plane near the (220) Brillouin zone using a series of constant-$\vec{Q}$ and constant-$\hbar\omega$ scans.  These scans probe TA and TO phonons polarized along [110] and propagating along [001], and they are directly comparable to those published on PbTiO$_3$~\cite{Shirane_1970} and PZN-8\%PT,~\cite{Gehring_waterfall} all of which were measured in the same scattering plane and in the same Brillouin zone.  The TA and TO phonon dispersion curves for PZT measured at 650\,K (60\,K above T$_{C1}$) are shown in Fig.~2a.  We find that the TA phonon branch in PZT is similar to, but slightly softer than, the corresponding TA phonon branch in PbTiO$_3$, which is shown in Fig.~2b.~\cite{Shirane_1970}  The TO phonon branch in PZT drops to very low energies at small reduced wave-vectors $\vec{q}=\vec{Q}-(2,2,0)$ and is similar in this regard to the soft TO phonon branch in PbTiO$_3$ measured at 783\,K (20\,K above $T_C$).  However, at large wave vectors the TO phonon branch in PZT is significantly softer than that in PbTiO$_3$ and closely resembles the TO phonon branch in the relaxor PZN-8\%PT, which is also shown in Fig.~2b.  In particular, the TO phonon branch in PZT appears to intercept the TA phonon branch at a non-zero wave vector near 0.05\,rlu (reciprocal lattice units), albeit this is substantially smaller than the corresponding wave vector (0.14\,rlu) in PZN-8\%PT.  This behavior suggests that the waterfall effect is present in PZT.

\begin{figure}[t]
\centering
\includegraphics[width=85mm]{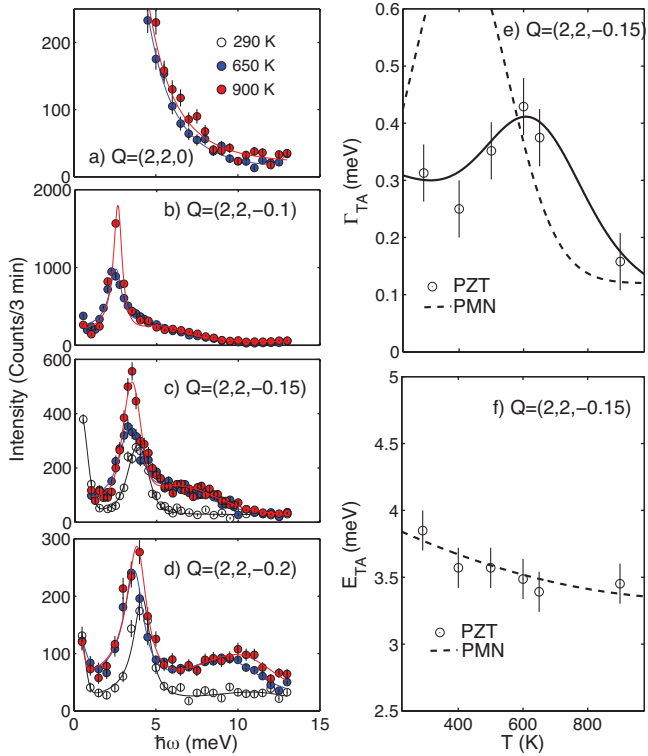}
\caption{Constant-$\vec{Q}$ scans measured at (a) (2,2,0), (b) (2,2,-0.1) , (c) (2,2,-0.15), and (d) (2,2,-0.2) at temperatures above and below $T_{C1} = 590$\,K.  Temperature dependencies of the TA phonon (e) linewidth $\Gamma_{TA}$ and (f) energy $E_{TA}$ for PZT (open circles) and PMN (dashed lines).  The solid line is a guide to the eye.}
\end{figure}

The nature of the waterfall effect in PZT is more fully revealed by an analysis of the constant-$\vec{Q}$ scans shown in Fig.~3A--3D, which were measured above and below the cubic-to-rhombohedral phase transition temperature $T_{C1}=590$\,K.  The TO mode at $\vec{Q}=(2,2,-0.2)$ is well-defined and softens on cooling from 900\,K to 650\,K, but it is nearly overdamped at the same temperatures at $\vec{Q}=(2,2,-0.1)$.  In fact for wave vectors $|\vec{q}| < 0.15$\,rlu the TO mode is so damped that it is not possible to locate the peak in a constant-$\vec{Q}$ scan.  This $\vec{q}$-dependent broadening of the soft TO mode is the hallmark of the waterfall effect and has been reported in various compositions of PMN-$x$PT and PZN-$x$PT.  The solid lines in these panels represent fits of the data to the sum of two uncoupled damped harmonic oscillators convolved with the instrumental resolution function.  The quality of the fits indicates that TA-TO mode coupling is weak in PZT for phonons propagating along [001].  A series of constant-$\vec{Q}$ scans were measured at $\vec{Q}=(2,2,-0.15)$ between 300\,K and 900\,K to characterize the temperature dependence of the TA phonon linewidth and energy.  Fig.~3E shows that the TA phonon linewidth in PZT broadens on cooling from above $T_{C1}$ and narrows slightly below.  This behavior is markedly different from that in PMN (dashed lines) for which the TA phonon linewidth broadens enormously on cooling.  The broadening in PMN has been shown to result from a coupling to a low-energy relaxational mode associated with the strong diffuse scattering.~\cite{Stock_2005,Stock_2012}  Assuming that a coupling of similar strength is present in PZT, then the comparatively weak temperature dependent TA phonon broadening would further support our finding of little or no diffuse scattering in this compound.  In Fig.~3f the TA phonon energy in PZT is shown to vary monotonically and changes little with temperature.  Indeed, the net shift in the PZT TA phonon energy over this broad temperature range is almost identical to that in PMN (dashed lines).~\cite{Stock_2005,Stock_2012}  We therefore exclude TA-TO mode coupling as the origin of the waterfall effect in PZT.  We also exclude REFs as the mechanism responsible for the waterfall effect.  At lower temperature (295\,K) only the TA mode is visible because the TO mode has hardened to energies that lie outside the range of these scans.  The softening of the TO mode on cooling from 900\,K to 650\,K and the subsequent hardening is consistent with the cubic-to-rhombohedral phase transition at $T_{C1}$ being driven by a soft, zone-center mode instability.~\cite{Phelan_PZT}

The lower-temperature phase transition for this composition of PZT at T$_{C2}=370$\,K corresponds to a doubling of the unit cell that is manifested by the appearance of new Bragg reflections at the R-points $(\frac{h}{2},\overline{\frac{h}{2}},\frac{l}{2})$, where $h$ and $l$ are odd and $h \neq l$.  This suggests that there may be a soft mode at the R-point.  In PMN, highly-damped (i.\ e.\ short-lived), soft zone-boundary modes are observed at both R-points and M-points (see Fig.~S3 in the SI Appendix), and structure-factor considerations preclude these from being octahedral tilt modes.  To learn if PZT exhibits similar dynamics, we measured constant-$\vec{Q}$ scans at the R-point $\vec{Q}=(\frac{3}{2},\frac{3}{2},-\frac{1}{2})$ at 290\,K, 650\,K and 900\,K, which are shown in Fig.~4a (top panel).  A broad distribution of scattering is observed in PZT below 10\,meV at 900\,K and 650\,K but is absent at 290\,K.  The shift in spectral weight from high to low energies on cooling from 900\,K is particularly apparent near $\hbar\omega=2.0$\,meV where an increase in scattering is seen at 290\,K.  To clarify this behavior we measured the scattering intensity at $\vec{Q}=(\frac{3}{2},\frac{3}{2},-\frac{1}{2})$ and $\hbar\omega=2.0$\,meV as a function of temperature.  The data (bottom panel) show that the scattering at 2\,meV increases on cooling from 900\,K to a maximum at near 450\,K and then decreases.  This behavior is precisely that expected for a soft mode: for $T > T_{C2}$ the soft mode is located above 2\,meV, which results in little or no scattering; but on cooling through $T_{C2}$ the soft mode drops to lower energies, thus moving into and then out of the instrumental window and causing the scattering intensity to first increase and then decrease.  This procedure mimics that employed by Swainson \emph{et al.} to demonstrate the softening of the R-point zone boundary mode in PMN.~\cite{Swainson_2009}

\begin{figure}[t]
\centering
\includegraphics[width=85mm]{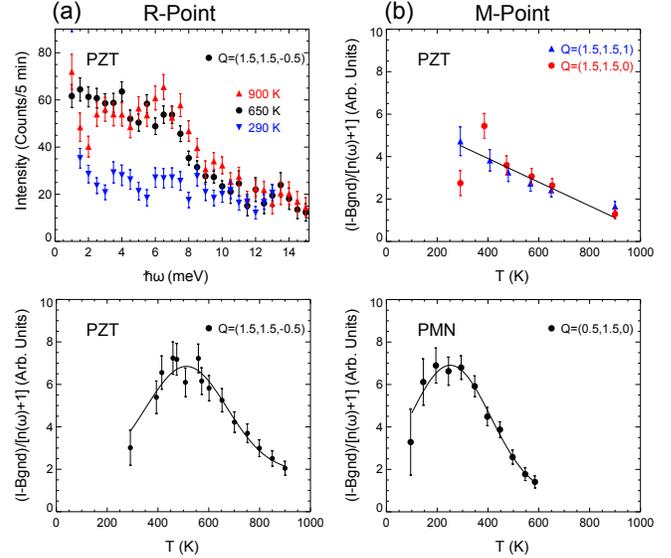}
\caption{(a) Upper panel - constant-$\vec{Q}$ scans measured at the R-point $(\frac{3}{2},\frac{3}{2},-\frac{1}{2})$ at 900\,K, 650\,K, and 290\,K.  Lower panel - background subtracted intensity, adjusted by the Bose factor $[n(\omega) + 1]$, measured at the same $\vec{Q}$ and at fixed $\hbar\omega=2.0$\,meV from 900\,K to 300\,K.  (b) Upper panel - neutron scattering intensity measured at the M-points $(\frac{3}{2},\frac{3}{2},0)$ (solid red circles) and $(\frac{3}{2},\frac{3}{2},1)$ (solid blue triangles) from 900\,K to 300\,K.  The background has been subtracted from the total scattering intensity and the result adjusted by the Bose factor $[n(\omega) + 1]$.  Lower panel - analogous data for the relaxor PMN measured by Swainson {\em et al.} (Ref. \cite{Swainson_2009}) at $\vec{Q}=(\frac{3}{2},\frac{3}{2},0)$.}
\end{figure}

We performed identical measurements on PZT at the M-points $(\frac{3}{2},\frac{3}{2},0)$ and $(\frac{3}{2},\frac{3}{2},1)$ at the same energy transfer $\hbar\omega=2.0$\,meV.  These data, shown in Fig.~4b (top panel), reveal completely different behavior as comparatively little temperature dependence is observed.  This is consistent with the fact that no superlattice reflections are seen at M-point locations in this composition of PZT.~\cite{Phelan_PZT} It also points to a fundamental difference in the dynamics between PZT and PMN for which concurrent, broad, soft zone-boundary modes were observed at \emph{both} R and M-points as indicated in the bottom panel.  At higher Ti concentrations ($x=0.475$) PZT exhibits a low-temperature monoclinic phase, and x-ray inelastic measurements have observed soft M-point modes.~\cite{Hlinka_PZT}  However no soft R-point mode was reported for this monoclinic composition.  We thus propose that a competition between ferroelectric and antiferroelectric fluctuations, as revealed by the simultaneous presence of heavily-damped (short-lived) soft modes at the R, M, and $\Gamma$ points, may represent a true dynamical signature of relaxor behavior.

\subsection{Ultrahigh piezoelectricity}
\ It is well known that pseudo-rhombohedral compositions of PMN-$x$PT and PZN-$x$PT located near the MPB exhibit enormous piezoelectric coefficients; the values of the longitudinal piezoelectric coefficient $d_{33}$ for single-crystal MPB compositions of PMN-$x$PT and PZN-$x$PT poled along [001] can exceed 2800\,pC/N and are the largest known.~\cite{Guo_2003,Cao_2004,Park}  The concepts of polarization rotation~\cite{Fu} and adaptive phases~\cite{Jin} were proposed to explain the ultrahigh electromechanical response in MPB systems like PZT and PMN-$x$PT.  More recently, Kutnjak \emph{et al.} reported that the origin of the giant electromechanical response at the MPB is related to the proximity of the system to a critical end point.~\cite{Kutnjak}  We note here that the $q$-integrated neutron elastic diffuse scattering intensity in PMN-$x$PT, which from the fluctuation-dissipation theorem is a measure of the static susceptibility, and $d_{33}$ share a similar composition dependence.~\cite{Matsuura_2006,Gehring_2012}  Both properties (in this case measured along [110]) peak near the MPB and then decrease precipitously in the tetragonal (Ti-rich) regime in which the relaxor phase is replaced by a conventional ferroelectric phase.~\cite{Stock_2006}  Given our finding that PZT exhibits essentially no such diffuse scattering one might expect that the corresponding value of $d_{33}$ in PZT should be significantly smaller than that in PMN-$x$PT.  We find that this is indeed true: the value of $d_{33}$ that we measured on a [001]-poled single crystal of PZT with a composition close to the MPB is 1,200\,pC/N.  Thus the optimal values of $d_{33}$ in single-crystal PMN-$x$PT and PZN-$x$PT are approximately 100\% greater than that in single-crystal PZT.  Insofar as PZT and PMN-$x$PT are identical systems save for the strength of the REFs, our results indicate that the underlying nanometer-scale, static, polar order in relaxors that is associated with the temperature-dependent diffuse scattering plays a seminal role in greatly amplifying the piezoelectric response beyond that in the non-relaxor PZT, which does not display an equivalent nanometer-scale polar order.  This conclusion is consistent with the theoretical results of Pirc \emph{ et al.} that suggest polar nanoregions play an important role in the electrostriction in relaxors.~\cite{Pirc_2004}  Kutnjak \emph{et al.} also showed that the phase transitions in relaxor-based ferroelectrics are first-order as an external electric field must be applied to reach the critical end point.  By contrast, PZT compositions near the MPB exhibit second-order phase transitions (critical points) at $T_C$ in zero field.~\cite{Kim_2012}  This suggests that the presence of REFs are at least as important to the large piezoelectric response as is the proximity of the system to a critical end point.  We note that this behavior is not without precedent.  In single crystals of highly magnetostrictive Galfenol, an alloy of magnetic Fe and non-magnetic Ga that is often likened to a magnetic relaxor and which should possess strong random magnetic fields, both the neutron elastic diffuse scattering intensity and the longitudinal magnetostrictive coefficient $\lambda_{100}$ increase dramatically as the system is doped towards a phase boundary and then decrease sharply as the boundary is crossed.~\cite{Cao_2009}

\section*{Conclusions}
We have performed a detailed comparison of the dielectric permittivity, nanoscale structure, and lattice dynamics of two highly similar, disordered, lead-oxide perovskite compounds that correspond to the limits of weak (PZT) and strong (PMN) random electric fields.  The relative weakness (or absence) of any elastic, anisotropic, and temperature-dependent diffuse scattering in PZT of the type observed in the relaxor PMN proves that the static, short-range polar displacements found in the lead-oxide, perovskite relaxors are not prevalent in ferroelectric PZT and therefore require the presence of strong REFs.  The lack of any frequency dispersion in the dielectric permittivity of PZT further indicates that the relaxor phase in these lead-oxide perovskites also requires the presence of strong REFs.  These conclusions have fundamental ramifications for the definition of the relaxor state because they imply that perovskite relaxors with homovalent $B$-site disorder such as BZT, and BST do not belong in the same classification as the lead-oxide perovskite relaxors, even though these systems all exhibit frequency-dependent dielectric susceptibilities.  Because the corresponding REFs in such homovalent compounds are weak in comparison to any crystalline anisotropy energy, the frequency-dependence of the dielectric permittivity must have a different physical origin.  Moreover, none of these homovalent compounds has been shown to exhibit the elastic, anisotropic, and temperature-dependent diffuse scattering seen in the lead-based relaxors,~\cite{Ge_2013} which we show boosts the piezoelectricity to record-setting levels.  In the cases of BZT and BST, as well as many other BaTiO$_3$-based ``relaxors," only temperature-independent, sheet-like diffuse scattering, similar to that seen in paraelectric BaTiO$_3$, has been observed using electron diffraction methods.~\cite{Liu_2007}  We hope that our results will motivate further experimental and theoretical studies of relaxors, which are essential to clarify the physics of random electric fields in condensed matter systems.

\begin{acknowledgments}
The authors acknowledge fruitful discussions with R.\ E.\ Erwin, D.\ K.\ Singh, and L.\ Kneller regarding the use and fabrication of a single-crystal silicon sample mount.  This work used facilities supported in part by the National Science Foundation under Agreement DMR-0944772.  The work at Simon Fraser University was supported by the US Office of Naval Research (Grants N00014-11-1-0552 and N00014-12-1-1045) and the Natural Science and Engineering Research Council of Canada. C.S. was partially supported by the Carnegie Trust. P.G. was supported by the Center for Nanophase Materials Sciences, which is sponsored at Oak Ridge National Laboratory by the Scientific User Facilities Division, Office of Basic Energy Sciences, US Department of Energy.
\end{acknowledgments}

\end{article}
\end{document}



\title{Supporting Information: The role of random electric fields in relaxors}

\author{Daniel Phelan\affil{1}{NIST Center for Neutron Research, National Institute of Standards and Technology, Gaithersburg, Maryland 20899, USA}, Christopher Stock\affil{1}{}\affil{2}{School of Physics and Astronomy, University of Edinburgh, Edinburgh, UK EH9 3JZ}, Jose A. Rodriguez-Rivera\affil{1}{}\affil{3}{Department of Materials Science and Engineering, University of Maryland, College Park, Maryland 20742, USA}, Songxue Chi\affil{1}{}\affil{3}{}, Juscelino Le\~{a}o\affil{1}{}, Xifa Long\affil{4}{Department of Chemistry and 4D LABS, Simon Fraser University, Burnaby, BC, V5A 1S6, Canada}, Yujuan Xie\affil{4}{}, Alexei A. Bokov\affil{4}{}, Zuo-Guang Ye\affil{4}{}, Panchapakesan Ganesh\affil{5}{Center for Nanophase Materials Sciences, Oak Ridge National Laboratory, Oak Ridge, Tennessee 37831-6494, USA} \and Peter M. Gehring\affil{1}{}\affil{6}{To Whom Correspondence Should Be Addressed (Email: peter.gehring@nist.gov)}}


\maketitle

\begin{article}

\section*{Materials and Methods}

\subsection*{Crystals}
Single crystals of Pb(Zr$_{1-x}$Ti$_x$)O$_3$ with composition $x=0.325$ were grown by a top-seeded solution growth technique. A platelet with ${100}$ faces and dimensions 1.1\,mm $\times$ 2.1\,mm $\times$ 3.3\,mm was prepared for this study; this is the same crystal on which we recently reported the results of neutron diffraction measurements.~\cite{Phelan_PZT}. The single crystal of Pb(Mg$_{1/3}$Nb$_{2/3}$)O$_3$ (PMN) used in the diffuse scattering measurement has ${100}$ faces and dimensions 10\,mm $\times$ 10\,mm $\times$ 10\,mm.

\subsection*{Dielectric and piezoelectric measurements}
The dielectric permittivities of PMN and Pb(Zr$_{0.675}$Ti$_{0.325}$)O$_3$ were measured as a function of temperature along [100] at various frequencies by means of a "Novocontrol" turnkey broadband dielectric spectrometer (Concept 20).  The piezoelectric coefficient $d_{33}$ for single-crystal Pb(Zr$_{0.54}$Ti$_{0.46}$)O$_3$ was calculated from the slope of the strain versus unipolar electric field curve measured using an MTI 2000 Fotonic sensor (Range 1).

\subsection*{Diffuse scattering measurements}
All neutron scattering measurements were performed on the BT7 and BT4 thermal-neutron triple-axis spectrometers and the cold-neutron Multi-Crystal Analyzer Spectrometer (MACS) located at the NIST Center for Neutron Research.  The incident and final neutron energies were selected via Bragg diffraction from the (002) reflections of a vertically-focusing pyrolytic graphite (PG) crystal assembly (BT7) and a flat PG analyzer crystal (BT4).  During the diffuse scattering measurements the incident and final neutron energies were fixed at 14.7\,meV on BT7 and BT4 and at 5.0\,meV on MACS, resulting in elastic energy resolutions of 1\,meV and 0.4\,meV, respectively, FWHM (full-width at half maximum).  All Miller indices are given with respect to a primitive cubic perovskite cell for which the lattice constant $a \approx 4.1$\,\AA.  Thus 1\,rlu (reciprocal lattice unit) $= 2\pi/a \approx 1.53$\AA$^{-1}$.

The 2D elastic scattering intensity maps for PMN and PZT, shown in Fig.~1C and D, \emph{Insets}, were measured on BT4 and MACS, respectively.  Data spanning l=1.11 to 1.15\,rlu have been symmetrized from those spanning l=0.89 to 0.85\,rlu for the case of PMN, and a powder ring has been removed from the intensity map for PZT.  Relaxor-like diffuse scattering contours are observed only in the PMN crystal.  Fig.~1C and D also show elastic scattering data (blue dots) obtained from identical scans. The orientation of these scans relative to the (001) Bragg peak is the same for each crystal and is indicated by the dashed blue arrow in each panel.  These scans were measured on the same instrument (BT7) at the same temperature (1.7\,K) using exactly the same instrumental configuration.  Thus the instrumental resolution is the same in both cases.  The resulting data were normalized to the same TA phonon for each crystal, which was measured under the same conditions.  Therefore these data provide a robust and quantitative upper bound on the strength of any diffuse scattering cross section that might be present in Pb(Zr$_{0.675}$Ti$_{0.325}$)O$_3$ relative to that in PMN.

Supplemental data on Pb(Zr$_{0.675}$Ti$_{0.325}$)O$_3$ are shown in Fig.~S1.  These data were measured in the $(HK0)$ scattering plane by scanning the wave vector $\vec{Q}$ along [100], [010], and [1-10] through the (100) Bragg peak.  All intensities have been normalized to monitor and are plotted on a logarithmic scale.  At 16\,K the scans along [100], [010], and [1-10] are all well-described by a single Gaussian function of $\vec{Q}$ plus a constant background, which is consistent with the absence of diffuse scattering.  At 620\,K the situation is the same, however a very small amount of Lorentzian character is observed for the scan measured along [010].  This Lorentzian component arises from critical fluctuations, which are visible given the proximity of the system to the second-order cubic-to-rhombohedral structural phase transition at $T_{C1} = 590$\,K.  These scans all confirm the absence of any measurable diffuse scattering cross section in this composition of PZT.

\begin{figure}[t]
\centering
\includegraphics[width=80mm]{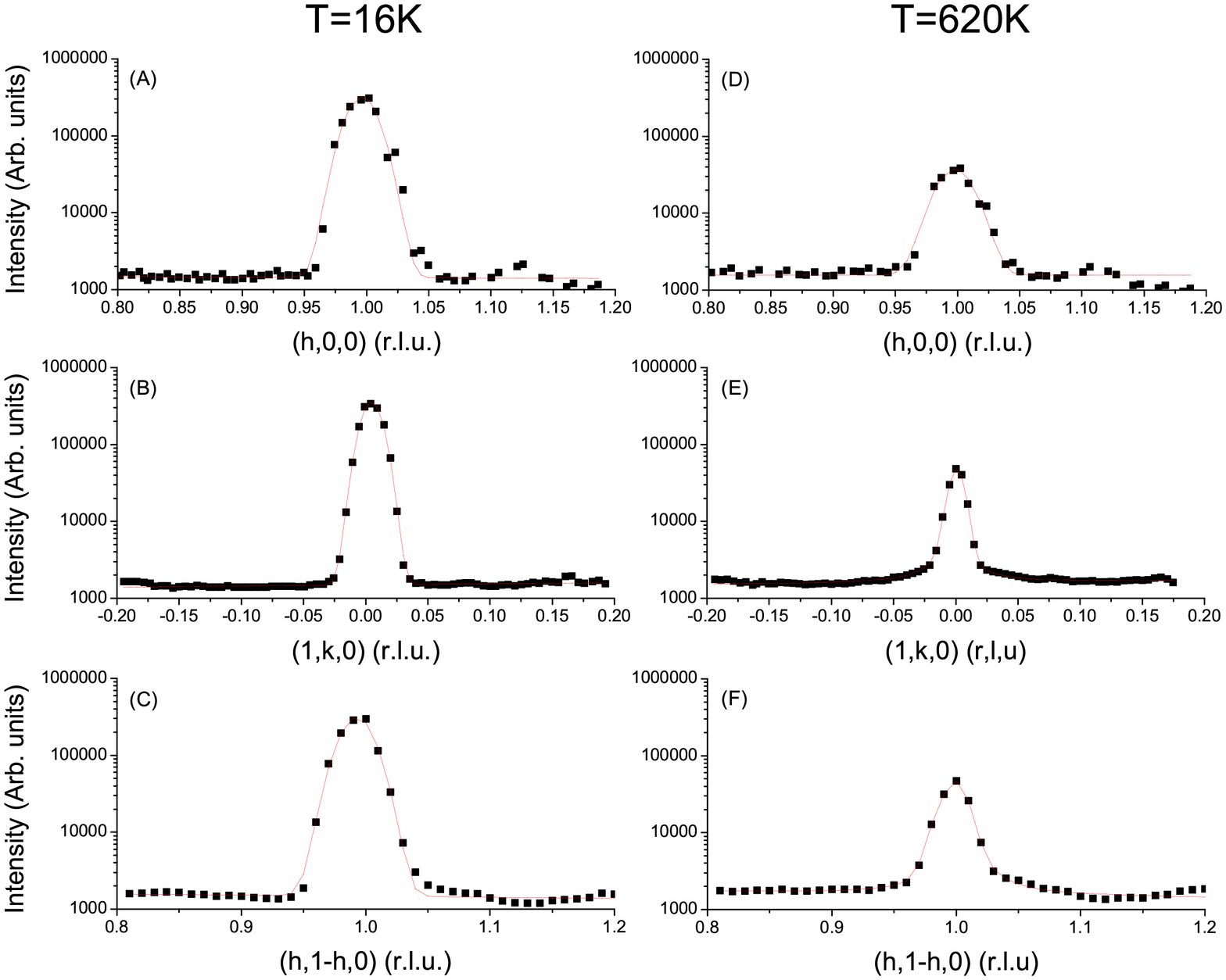}
\caption{Elastic scattering data measured on single-crystal PZT ($x=0.325$) at 16\,K and 620\,K in the $(HK0)$ scattering plane using the cold-neutron MACS spectrometer, which provides an elastic energy resolution of 0.4\,meV (FWHM).  Data are plotted on a logarithmic intensity scale.}
\end{figure}

\subsection*{Heuristic REF Model} Given that strong REFs are present in Pb-based perovskites that show a relaxor phase, we develop a heuristic model to understand the role played by REFs as well as the order parameter in these materials.  We assume that the system can be divided into regions of weak and strong REFs as indicated schematically in Fig. S2, which is true for Pb(Sc$_{1/2}$Nb$_{1/2}$)O$_3$,~\cite{Ganesh_2010} and we write down a general correlation function of the random variables $\chi (r)$  with translational symmetry $C(r,r') = C(r-r')$ as a sum over these regions:

\begin{equation}\label{Corr}
C(r,r') = \sum C^{11}(r,r') + \sum C^{22}(r,r') + \sum C^{12} (r,r').
\end{equation}

The superscripts 1,2 in each term refer to correlations between the random variable at position $r$ in region 1 or 2 with the same variable at position $r'$ in region 1 or 2.  We define region 1 to have weak local REFs and region 2 to have strong local REFs.  Given that, by definition, the local REFs are weak in region 1, the local polar displacements will tend to be larger than those in region 2 because the damping of the polar soft-mode will be weaker.  Therefore the first term in Eq.~1 is dominant, and it predominantly gives the diffuse scattering as shown in Ref.~\cite{Ganesh_2010}. We further expand this term as follows:

\begin{equation}\label{Corrapprox}
C(r,r') \sim \sum C^{11} (r,r') = \sum C_a^{11} (r,r') + \sum C_b^{11} (r,r').
\end{equation}

The subscript $a$ refers to correlation terms where both $r$ and $r'$ are in the same region-1 and the subscript $b$ refers to the terms in two separate regions of type 1, i.\ e.\ they are spatially separated by a region-2 with strong REFs (see Fig. S2).  As such $C_b$ is more highly screened than $C_a$, and the strength of the screening should depend on the strength of the local electric field in region-2, which we denote by $|h_2(r)|$.  As a first approximation we assume a position-independent local field in region-2 so that we may approximate $C_b^{11}(r,r') \sim e^{-f(|r-r'|)/|h_2|} C_a^{11} (r,r')$, in a similar spirit to the screened Yukawa potential applied to a Coulomb term in electrostatics.  Here $f$ is a short-ranged function independent of region-2 that depends only on the spatial distance, but for which the screening length is governed by the local field strength $|h_2|$.  This leads to an expression of the form:

\begin{equation}\label{CorrYukawa}
C(r,r') \sim \sum C_a^{11} (r,r') + \sum e^{-f(|r-r'|)/|h_2|} C_a^{11} (r,r').
\end{equation}

Eq.~3 illustrates the role of the local field strength in region 2, $|h_2|$.  In a ferroelectric system where $|h_2| \rightarrow 0$, the correlations inside a particular region 1 will dominate the correlation function.  As region 1 is relatively ordered (due to the weaker REFs), one essentially recovers long-range order (resulting in a Bragg peak). In a relaxor like PMN, where the strength of $|h_2| \gg 0$, the exponential pre-factor in Eq.~3 is ideally unity due to overdamped conditions and the system develops more structure at longer wavelengths, i.e. shorter wave vectors $q$, leading to diffuse scattering.  For intermediate values of the field strength this contribution will be modulated.  This heuristic model then suggests that the strength of the local REFs governs the appearance of diffuse scattering seen only in the relaxor phase.  Within a simple, damped harmonic model the local REF strength can be related to the dispersion of the dielectric constant by setting the random force equal to $Zeh_2$, where $Ze$ is the local charge of the ions. This suggests that the frequency dispersion of the permittivity and the diffuse scattering have the same microscopic origin, i.\ e.\ a region of strong REFs percolating throughout an otherwise weak-REF region. As such, the state of the overall system can be parameterized by the strength of the local electric field in the strong REF region, which is also roughly equal to the average local electric-field of the system.

\begin{figure}[t]
\centering
\includegraphics[width=70mm]{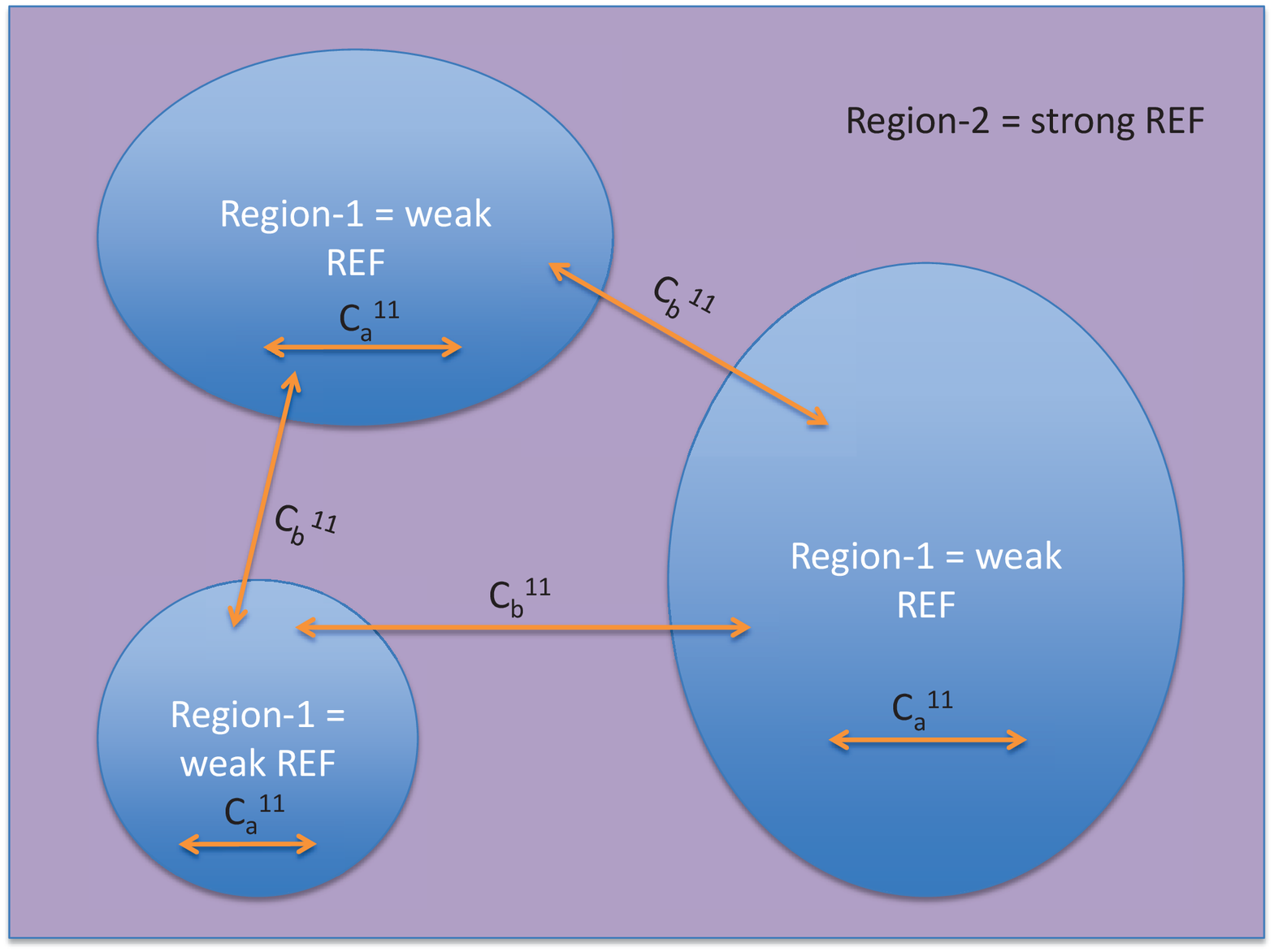}
\caption{Schematic representation of the regions of weak (1) and strong (2) REFs and the associated correlation functions.}
\end{figure}

\subsection*{Phonon measurements}
All phonon data were measured on BT7 for which the initial neutron energy was varied while the scattered energy was held fixed at 14.7\,meV; two highly-oriented PG filters were placed in the scattered beam to suppress higher-order contaminant neutron wavelengths.  The resulting elastic energy resolution was 1~meV FHWM.  A standard correction was applied to the data from constant-$\vec{Q}$ scans to compensate for the presence of higher harmonics in the incident neutron beam monitor.  A special single crystal silicon sample mount was fashioned to hold the PZT crystal; this was done to minimize the background scattering associated with the sample holder.  All phonon data in this study were obtained in the $(HHL)$ scattering plane.

Fig.~S3 shows extensive phonon data, replotted from ref.~3, that were measured on a different PMN single crystal.  These data, also obtained in the $(HHL)$ scattering plane, clearly reveal the presence of broad zone-boundary modes at a variety of different R and M-points, which were shown by Swainson \emph{et al.} to soften on cooling from 600 to 300\,K.~\cite{Swainson_2009}

\begin{figure}[b]
\centering
\includegraphics[width=80mm]{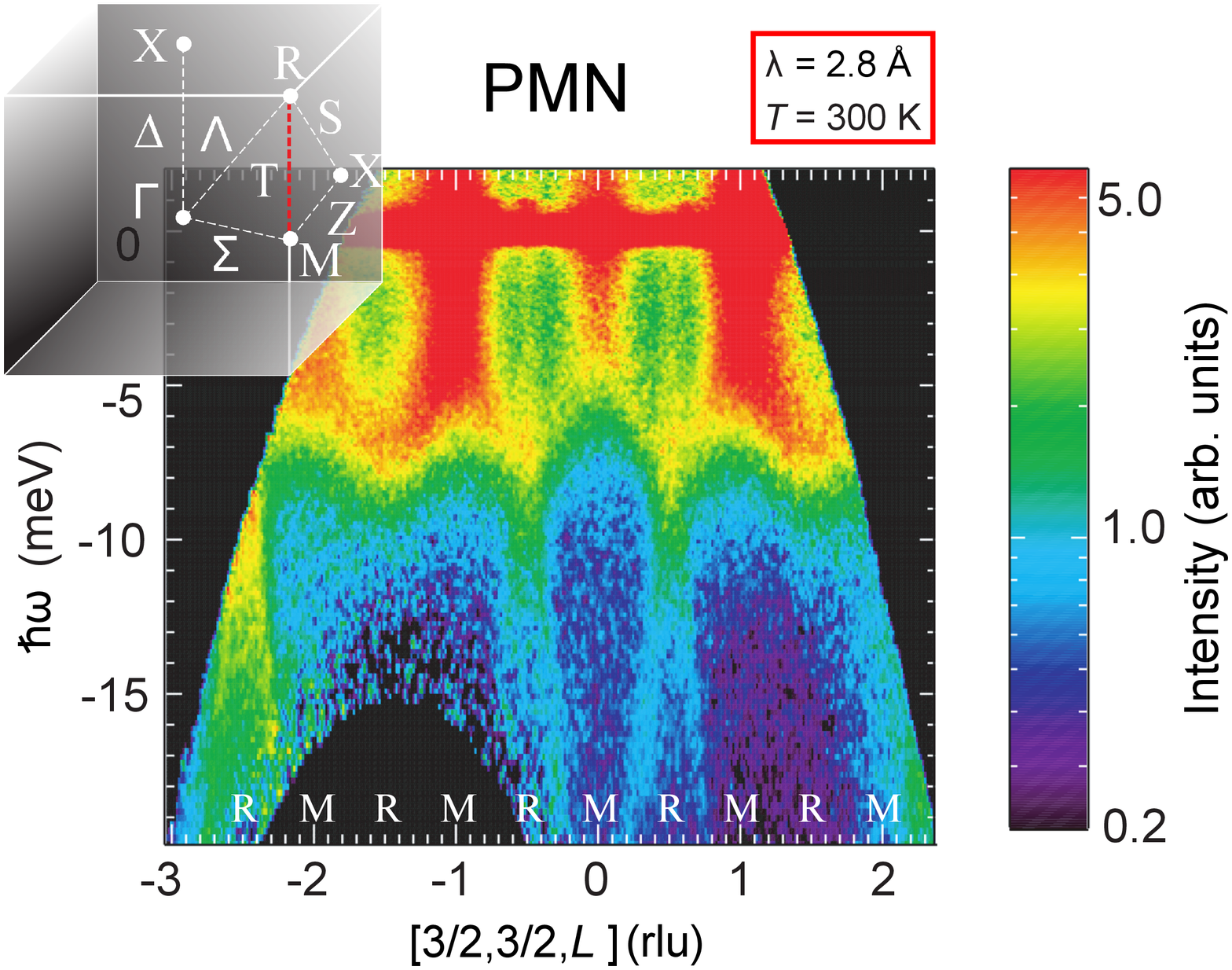}
\caption{Neutron inelastic scattering (log-scale) measured in the $(HHL)$ scattering plane at 300\,K showing the presence of soft zone-boundary modes located at various R (weak) and M-points (strong) in PMN (data have been replotted from Swainson \emph{et al.} \cite{Swainson_2009}).  Locations of the R and M-points are indicated in the Brillouin zone (grey cube) as well as the line "T" (dashed red line) that connects them.}
\end{figure}

\section*{Figure Legends}

\noindent \textbf{Fig. S1.} Elastic scattering data measured on single-crystal PZT ($x=0.325$) at 16 and 620\,K in the $(HK0)$ scattering plane using the cold-neutron MACS spectrometer, which provides an elastic energy resolution of 0.4\,meV (FWHM).  Data are plotted on a logarithmic intensity scale.
\\
\\
\noindent \textbf{Fig. S2.} Schematic representation of the regions of weak (region 1) and strong (region 2) REFs and the associated correlation functions.
\\
\\
\noindent \textbf{Fig. S3.} Neutron inelastic scattering (log-scale) measured in the $(HHL)$ scattering plane at 300\,K showing the presence of soft zone-boundary modes located at various R (weak) and M-points (strong) in PMN (data replotted from Swainson \emph{et al.} \cite{Swainson_2009}).  Locations of the R and M-points are indicated in the Brillouin zone (grey cube) as well as the line "T" (dashed red line) that connects them.

\end{article}